\input harvmac.tex
\def\exp{{\rm exp}}

\def\frac#1#2{{#1\over#2}}

\lref\AlZ{Zamolodchikov, Al.B.: unpublished}

\lref\ZamAl{
Zamolodchikov, Al.B.:
Two-Point correlation function
in Scaling Lee-Yang model.
Nucl. Phys. {\bf B348}, 619-641 (1991)}

\lref\ABZ{Zamolodchikov, A.B.:
Integrable field theory from conformal field theory.
Adv. Stud. in Pure Math.
{\bf 19}, 641-674 (1989)}

\lref\Zarn{Zamolodchikov, Al.B.: Mass scale
in the Sine-Gordon model and its
reductions. Int. J. Mod. Phys. {\bf A10}, 1125-1150  (1995) }

\lref\Coulmen{Coleman, S.: The quantum sine-Gordon
equation as the Massive Thirring model. Phys. Rev. {\bf D11},
2088-2097 (1975)}

\lref\ZaZa{Zamolodchikov, A.B. and Zamolodchikov, Al.B.:
Factorized S-matrices in two dimensions as the exact
solutions of certain relativistic quantum field theory models.
Ann. Phys. (N.Y.) {\bf 120}, 253-291 (1979) }

\lref\Luk{Lukyanov, S.: Free Field Representation for Massive
Integrable Models,
Commun. Math. Phys. {\bf 167}, 1, 183-226 (1995)\semi
Lukyanov, S.: Correlators of the Jost
functions  in the Sine-Gordon Model. Phys. Lett.
{\bf B325}, 409-417 (1994)}

\lref\deV{ Destri, C. and  de Vega, H.:
New exact result in affine Toda
field theories: Free energy and
wave functional renormalization.
Nucl. Phys., {\bf B358}, 251-294 (1991)}

\lref\BerK{Berg, B., Karowski, M. and
Weisz, P.: Construction of Green's functions from
an exact S-matrix. Phys. Rev. {\bf D19}, 2477-2479 (1979)}  

\lref\Leclair{LeClair, A.: Restricted Sine-Gordon theory
and the minimal 
conformal series. Phys. Lett. {\bf B230}, 103-107 (1989)} 

\lref\Fedya{Smirnov, F.A.: Form-factors in completely
integrable models of
quantum field theory. Singapore: World Scientific 1992}

\lref\Sm{Smirnov, F.A.: Reductions of Quantum Sine-Gordon Model
as Perturbations of Minimal Models of Conformal Field Theory.
Nucl. Phys. {\bf B337}, 156-180 (1990)}

\lref\KouM{Koubek, A. and Mussardo, G.: 
On the operator Content of the Sinh-Gordon Model.
Phys. Lett. {\bf B311}, 193-201 (1993)}

\lref\MiW{Jimbo, M. and Miwa, T.: Quantum KZ equation
with\ $|q|=1$\ and correlation functions
of  the XXZ model in the gapless regime.
J. Phys. {\bf A29}, 2923-2958 (1996)}

\lref\luth{Luther, A.: Eigenvalue
spectrum of interacting massive fermions in one dimension.
Phys. Rev. {\bf B14}, 5, 2153-2159 (1976)}

\lref\Baxter{Baxter, R.J.: Exactly Solved Models
in Statistical Mechanics. London: Academic Press 1982}

\lref\Jons{Johnson, J.D., Krinsky, S. and McCoy, B.M.:
Vertical-arrow correlation length in
the eight-vertex model and the low-lying excitation
of the X-Y-Z Hamiltonian.
Phys. Rev. {\bf A8}, 2526-2547 (1973)}

\lref\BaxKel{Baxter, R.J. and Kelland, S.B.:
Spontaneous polarization of
the eight-vertex model.
J. Phys. C: Solid State Phys., {\bf 7}, L403-L406 (1974)}

\lref\MiwJ{Sato, M., Miwa, T. and Jimbo, M.: Holonomic Quantum
Fields. IV. Publ. Rims, Kyoto Univ. {\bf 15}, 871-972 (1979)}

\lref\BerLec{Bernard, D. and Leclair, A.: Differential
equations for sine-Gordon correlation functions
at the free fermion point. Nucl. Phys. {\bf B426},
543-558 (1994)}

\lref\WuM{Wu, T.T., McCoy, B.M., Tracy, C.A. and
Barouch, E.: Spin-spin correlation functions for the
two-dimensional Ising model: Exact theory in
the scaling region. Phys. Rev.  {\bf B 13}, 1, 316-374
(1976)}

\Title{\vbox{\baselineskip12pt\hbox{CLNS 96/1444}
\hbox{RU-96-107}                
\hbox{hep-th/9611238}}}
{\vbox{\centerline{
Exact expectation values of local fields}
\vskip6pt
\centerline{in quantum sine-Gordon model}}}

\centerline{Sergei Lukyanov$^{1,3}$ \footnote{$^{*}$}
{e-mail address: sergei@hepth.cornell.edu} 
and Alexander Zamolodchikov$^{2,3}$}
\centerline{$^1$Newman Laboratory, Cornell University}
\centerline{ Ithaca, NY 14853-5001, USA}
\centerline{$^2 $Department of Physics and Astronomy,
Rutgers University}
\centerline{ Piscataway,
NJ 08855-0849, USA}
\centerline{and}
\centerline{$^3$L.D. Landau Institute for Theoretical Physics,}
\centerline{Chernogolovka, 142432, RUSSIA}

\centerline{}
\centerline{}
\centerline{}

\centerline{\bf Abstract}

We propose an explicit expression for vacuum expectation values 
$\langle\,  e^{ia\varphi}\, \rangle$ of the
exponential fields in the  sine-Gordon model. Our expression
agrees both with semi-classical results in the
sine-Gordon theory and with
perturbative calculations in the  Massive Thirring model. We use this
expression to make new predictions about the large-distance asymptotic
form of the two-point
correlation function in the  XXZ spin chain.

\Date{November, 96}
\vfill

\eject

Many $2D$ Quantum 
Field Theories (QFT) can be realized as Conformal Field
Theories (CFT) perturbed by some relevant operators \ABZ.
In these cases the
correlation function of
the local fields ${\cal O}_a (x)$ of the perturbed
theory
\eqn\cf{\langle\, {\cal O}_{a_1}(x_1)...{\cal O}_{a_N}(x_N)
\, \rangle }
are defined in a formal way through the Conformal Perturbation Theory
(CPT) as
\eqn\cpt{Z(\mu)^{-1}\ \langle\,
{\cal O}_{a_1}(x_1)...{\cal O}_{a_N}(x_N)\ 
e^{-\mu\int d^2 x\, \Phi(x)}\, \rangle_{CFT}\, ,}
where 
$$Z(\mu) = \langle\, e^{-\mu\int d^2 x\, \Phi(x)}\,
\rangle_{CFT}\, ,$$
$\Phi$ is the local field taken as the perturbation, $\mu$
is the CPT expansion parameter and \hbox{$\langle\,...
\, \rangle_{CFT}$} denote the
expectation values in the original CFT. Of course,
the expression\ \cpt\  
for the correlation functions\ \cf\ can 
not be taken literally. As is
well known, the power series
expansion in $\mu$ generated by\ \cpt\ 
is always plagued by infrared divergences,
reflecting the fact that
the correlation functions
\ \cf\  contain non-analytic in $\mu$ terms.
However these  non-analytic
terms are rather the property of the vacuum
$|\, 0\, \rangle$
in $\langle\,...\, \rangle =
\langle\, 0\, |...|\, 0\, \rangle/\langle\, 0\, |\, 0\, \rangle$
then of the local operators ${\cal O}_a$. 
The local fields are believed to satisfy the OPE algebra
\eqn\ope{{\cal O}_{a}(x)\, {\cal O}_{b}(y)\simeq
\sum_{c}\ C_{ab}
^{ \,c}(x-y)\ {\cal O}_{c}(y)\ .}
By the very virtue
of these relations the c-number coefficient functions
$C$
are not very sensitive to the long-range properties of the theory
(in particular, the CPT for the coefficients $C$ do not contain any
infrared divergences) and therefore one expects that they can be
expanded into a power series
in $\mu$ with finite radius of convergence. 
On the other hand, by
successive application
of\ \ope\ any correlation function\ \cf\ can be
reduced down to
the one-point functions $ \langle\, 
{\cal O}_{a}(x)\,  \rangle$. Simple
dimensional analysis shows that these quantities are non-analytic in
$\mu$
\eqn\opf{
\langle\,{\cal O}_{a}(x)
\, \rangle= G_a \  \mu^{{\Delta_a}\over{1-\Delta}}\ ,}
where $\Delta_a$ are
conformal dimensions of the fields ${\cal O}_a$ and
$G_a$ are constants. In writing\ \opf\ 
we have assumed that the
perturbing operator $\Phi$ consists 
of a single scaling field of the
conformal dimension $\Delta$. In this sense it is the one-point
functions\ \opf\  that contain
all the non-perturbative information about the 
QFT which can not be extracted directly from CPT\ \ZamAl.
Therefore the problem
of calculation of the one-point
functions has fundamental significance
\foot{Although our discussion explicitly concerns $2D$
QFT, the  general statements above are expected to hold in
multidimensional QFT as well.}.

During the last two decades much progress
has been made in understanding
$2D$ Integrable QFT. However the closed expressions for the one-point
functions are
obtained in very few cases. Most significant result
obtained in this direction
so far was exact computation in many cases the
expectation value of the perturbing operator $\Phi$
in\ \cpt\  with the
help of Thermodynamic Bethe
Ansatz and similar techniques \Zarn. In this 
paper we
propose an explicit expression for the one-point functions of all 
exponential fields ${\cal O}_a (x) =
e^{ia\varphi}(x)$ in the  sine-Gordon QFT.

The sine-Gordon model is defined by the Euclidean  action
\eqn\sg{
{\cal A}_{SG} =
\int\ d^2 x\ \biggl\{\ 
{1\over{16\pi}}(\partial_\nu \varphi)^2
-2\mu\, \cos( \beta\varphi ) \ \biggr\}\ ,}
where $\beta$ and $\mu$ are parameters \foot{Note that our
notation $\beta$ for the sine-Gordon coupling constant differs by a 
factor $\sqrt{8\pi}$ from conventional one
\Coulmen.}. The renormalization required
with this definition is relatively simple.
The constant
$\beta$ does not
renormalize while $\mu$ is renormalized multiplicatively.
In order to give
the constant $\mu$ a precise meaning one has to fix the
normalization of the field 
\ $\cos(\beta\varphi)$. A convenient normalization 
which we will accept here corresponds to the short distance limit of
the two-point function
\eqn\coscos{
\langle\,
\cos(\beta\varphi)(x)\, \cos(\beta\varphi)(y) \, \rangle
\to{1\over 2}\ 
|x-y|^{-4\beta^2} \quad {\rm as} \quad |x-y|\to 0\ .}
This normalization is the most
natural one if one understands\ \sg\ 
as a 
perturbed CFT, with the field $\cos(\beta\varphi)$ taken as the
perturbing operator. Under this normalization the field
$\cos(\beta\varphi)$ has the
dimension $\big[\, length\, \big]^{-2\beta^2}$
and correspondingly the dimension of
the constant $\mu$ is $\big[\, length\, \big]^{2\beta^2-2}$.

The theory\ \sg\  
has a discrete symmetry $\varphi \to \varphi + 2\pi
n/\beta$ with any integer $n$.
In this paper we study\ \sg\ in the 
domain $0<\beta^2 < 1$,
where the above symmetry
is spontaneously broken (and 
the QFT\ \sg\ is massive),
so that the theory has infinitely many
ground states $|\,  0_n\, \rangle $
characterized by the associated expectation values
of the field $\varphi$, $\langle\,
\varphi\, \rangle_n = 2\pi n/\beta$, 
where $\langle\, 
\ldots\, \rangle_n =
\langle\, 0_n\, |\ldots|\, 0_n
\, \rangle/\langle\,0_n\,  |
\, 0_n\, \rangle$. In what follows we concentrate our 
attention on one of these ground states, say
$|\, 0_0\, \rangle $, and use the 
notation $\langle\, \dots\, \rangle \equiv
\langle\, \dots\, \rangle_0$.

The sine-Gordon model
admits an equivalent description as the Massive 
Thirring model\ \Coulmen
\eqn\mtm{
{\cal A}_{MT} = \int\,  d^2 x\ 
\biggl\{\ 
i\, {\bar\psi}\gamma^{\nu}\partial_{\nu}\psi
-{\cal M}\, {\bar\psi}\psi - {g\over 2}\,  \big({\bar\psi}
\gamma^{\nu}\psi\big)^2\ \biggr\}\ ,}
where $\psi, {\bar\psi}$ is 
the Dirac field, the four-fermion coupling
constant $g$ relates to $\beta$ in\  \sg\  as
\eqn\betag{
{g\over\pi}= {1\over{2\beta^2}}-1\ ,}
and ${\cal M}$ is the
mass parameter.  
In particular, the fermion current of\ \mtm\ is related to the 
field $\varphi$ in\ \sg\ as
\eqn\jphi{j^{\nu} \equiv {\bar\psi}\gamma^{\nu}\psi =
-{\beta\over{2\pi}}\  \epsilon^{\nu \lambda }\,
\partial_{\lambda}\varphi\  .}

The QFT\ \sg\ is integrable and its on-shell solution,
i.e. spectrum of
particles and S-matrix, is well known \ZaZa.
The theory contains soliton $S$,
antisoliton ${\bar S}$
(these particles 
coincide with the ``fundamental
fermions'' of the Lagrangian
\mtm)  and a number of soliton-antisoliton 
bound states (``breathers'') $B_n$, $n = 1, 2, \ldots, < 1/\xi$; 
here and below the notation  
\eqn\xibeta{\xi = {{\beta^2}\over{1-\beta^2}} }
is used. The lightest of these bound states\ $B_1$
coincides with the 
particle associated
with the field $\varphi$
in perturbative treatment 
of the QFT\ \sg. Its mass $m$ is given by
\eqn\mM{m = 2M\, \sin\big( \pi\xi/2\big)\ ,} 
where $M$ is the soliton mass.
Exact relation between the particle 
masses and the parameter
$\mu$ in\ \sg\ 
defined with respect to the 
normalization\ \coscos\ was 
found in\ \Zarn
\eqn\muM{\mu = {{\Gamma(\beta^2)}\over{\pi\, \Gamma(1-\beta^2)}}
\biggl[\, M\ {{\sqrt{\pi}\,\Gamma\big({1\over 2}+{\xi\over 2}\big)}
\over{2\,\Gamma\big({\xi\over 2}\big)}}\,  \biggr]^{2-2\beta^2}\ . }
This allows one to derived the vacuum expectation value of the
field $\exp(i\beta\varphi)$ using an obvious relation
\eqn\ebetaphi{
\langle\, 
e^{i\beta\varphi}\, \rangle =-{1\over 2}\ \partial_{\mu}
f(\mu)\ ,} 
where $f(\mu)$ is the specific free energy
$f(\mu)=-{1\over V}\log Z$,
which is also known exactly\ \Baxter,\ \deV
\eqn\fbulk{f(\mu)= -{{M^2}\over 4}\,\tan\big( \pi\xi/ 2 \big)\  .}
Combining
\ \mM-\fbulk\    one finds
\eqn\ebetaphi{
\langle\, e^{i\beta\varphi}
\, \rangle = {{(1+\xi)\,\pi\,\Gamma(1-\beta^2)}
\over{16\,\sin(\pi\xi)\,\Gamma(\beta^2)}}\ 
\biggl(\, {{\Gamma({1\over 2}+{\xi\over 2})\,\Gamma(1-{\xi\over 2})}
\over{4\,\sqrt{\pi}}}
\, \biggr)^{2\beta^2-2}\ m^{2\beta^2}\ ,}
where we have assumed that the field $e^{i\beta\varphi}$ is normalized
according to\ \coscos.

In this paper we study the expectation
values\ \opf\  of the exponential
fields ${\cal O}_a (x) = e^{ia\varphi}(x)$ 
in the sine-Gordon model.
Like in\ \coscos, we assume that these fields 
are normalized in accordance
with the following short distance limiting
form of their two-point functions
\eqn\norm{
\langle\,e^{ia\varphi}(x)\, 
e^{-ia\varphi}(y)\, \rangle \to |x-y|^{-4a^2}\qquad
{\rm as} \qquad |x-y|\to 0\ .}
Under this
normalization the field
$e^{ia\varphi}$ has the dimension
$\big[\, length\, \big]^{-2a^2}$.

For two special values of $\beta^2$ the one-point function
\eqn\expect{{\cal G}_a=\langle\,e^{ia\varphi}\, \rangle }
admits direct calculation.

a) The semi-classical limit $\beta^2\to 0$. In Appendix A
we study the expectation values  \expect\ 
with $a = \alpha/\beta$ for
$\beta^2\to 0$ and fixed $\alpha$, and obtain the result
\eqn\expcl{
\langle\, e^{i{\alpha\over\beta}\varphi} \, \rangle\to D(\alpha)
\ \bigg({m\over 4}\bigg)^{2{{\alpha^2}\over{\beta^2}}}\ 
\exp\biggl\lbrace{1\over\beta^2}\int_{0}^{\infty}{{dt}\over t}\, 
\bigg[ \ {{\sinh^2 (2\alpha t)}\over{t\,\sinh (2t)}}-2\alpha^2\,
e^{-2t}\ \bigg]\, 
\biggl\rbrace\ ,}
where $D(\alpha)$ can be calculated
as the functional determinant (A.11).

b) $\beta^2 = 1/2$. In this case the Thirring
coupling constant\ \betag\ 
vanishes and the sine-Gordon model is equivalent to the free-fermion
theory. In
Appendix B we use this simplification to obtain for $\beta^2=1/2$ and
$\big|\, \Re e\, a\, \big| < 1/\sqrt{2}$
\eqn\expff{
\langle\, e^{ia\varphi}\, \rangle = \bigg({M\over
2}\bigg)^{2a^2}\exp\biggl\lbrace\int_{0}^{\infty}\, {{dt}\over t}
\ \biggl[\ {{\sinh^2 (\sqrt{2}at)}
\over{\sinh^2 (t)}}-2a^2\,e^{-2t}\ \biggr]\,  
\biggr\rbrace\ ,}
where $M$ is the mass of the free fermion field.

Evident similarity between\ \expcl\ and\ \expff\ suggests the 
following expression for the expectation
value\ \expect\ for generic 
$\beta^2 < 1$ and $\big|\, \Re e\, a\, \big| <1/(2\beta) $ 
\eqn\main{\eqalign{
{\cal G}_a=&
\bigg[\,
{{m\,\Gamma\big({1\over 2}+
{\xi\over 2}\big)\Gamma\big(1-{\xi\over 2}\big)}
\over{4\sqrt{\pi}}}\, \bigg]^{2a^2}
\times\cr
&\exp\biggl\lbrace\int_{0}^{\infty}{{dt}\over t}
\bigg[\ {{\sinh^2 ( 2a\beta t )}\over{2\sinh(\beta^2 t)\, \sinh(t)\, 
\cosh\big((1-\beta^2)t\big)}}-
2a^2\,e^{-2t}\ \bigg]\, \biggl\rbrace\ .}}
The pre-exponential factor in\ \main\ 
is chosen to agree with\ \ebetaphi.
The formula\ \main\  is our conjecture for\ \expect.

The expression\ \main\ 
can be expanded in power series in $\beta^2$ or  
in\ $g$\ \betag\ and 
the coefficients of these expansions can be 
compared with the perturbative
calculations for the actions\ \sg\ and\ \mtm, respectively.
To perform
this check it is more convenient to consider the expectation values
$\langle\,
\varphi^{2n}\, \rangle$
which can be readily obtained by expanding\ \expect\ 
in power series in $a^2$. Let us define ``fully connected'' one-point
functions $\sigma_{2n}$
as
\eqn\sigmadef{\langle\, e^{ia\varphi}\, \rangle\equiv
1+\sum_{n=1}^{\infty}{{(-a^2)^n}
\over{(2n)!}}\ \langle\, \varphi^{2n}\, \rangle= 
\exp\bigg(\sum_{n=1}^{\infty}{{(-a^2)^n}
\over{(2n)!}}\ \sigma_{2n}\bigg)\ ,}
so that
\eqn\sigmas{
\sigma_2 = \langle\,\varphi^2\, \rangle\, ;
\qquad \sigma_4 = \langle\,\varphi^4 \, \rangle
-3\, \langle\,\varphi^2\, \rangle^2\ ;
\qquad\ldots\ .}
Expansion of \ \main\  gives
\eqn\sigman{\eqalign{
&\sigma_{2}=-4 \, {\rm log}\,\biggl\{\, 
{{m\,\Gamma\big({1\over 2}+{\xi\over 2}\big)
\Gamma\big(1-{\xi\over 2}\big)}
\over{4\sqrt{\pi}}}\, \biggr\}-\cr
&\ \ \ \ \ \ 4\, \int_{0}^{\infty}\, {dt\over t}\  \biggl\{\, 
{{\beta^2\, t^2}
\over{\sinh(\beta^2 t)\, \sinh(t)\, \cosh\big((1-\beta^2)t \big)}}-
e^{-2 t}\, \biggr\}\ ,\cr
&\sigma_{2n}=(-1)^n\, 
4^{2n-1}\, \beta^{2 n} \ \int_{0}^{\infty}
{{t^{2n-1}\,dt}\over{\sinh(\beta^2 t)\, 
\sinh(t)\, \cosh\big((1-\beta^2)t\big)}}\ , \ \ \ \ n>1\ .}}
We show in Appendix C  that $\sigma_2$
and\ $\sigma_4$\ in\ \sigman\ agree 
with the perturbation theory
for\ \sg\ up to $\beta^4$, and that\ $\sigma_2$\ 
agrees with the perturbation
theory for\ \mtm\ up to $g$.

Clearly, more checks of\ \main\ are desirable.
We note in this connection
that the expectation value\ \expect\ 
controls both short and long
distance asymptotics of the two-point correlation function 
\eqn\twopt{
{\cal G}_{a a'}\big(|x-y|\big)=
\langle\, e^{ia\varphi}(x)\, e^{ia'\varphi}(y)\, \rangle }
with $|a+a'|<\beta/2$. Indeed, if this
inequality is satisfied the short
distance limit of\ \twopt\ is dominated by OPE
\eqn\ope{e^{ia\varphi}(x)\, e^{ia'\varphi}(y)\to
\big|x-y\big|^{4aa'}\ e^{i(a+a')\varphi}(y)
\qquad {\rm as} \qquad |x-y|\to 0\  .}
Therefore
\eqn\twoass{
{\cal G}_{aa'}(r)\quad\to\quad
\Biggl\lbrace{{{\cal G}_{a+a'}\ r^{4aa'}
\qquad {\rm as} \qquad r\to 0}
\atop{\ \ \, {\cal G}_{a}\ {\cal G}_{a'}\qquad\qquad
{\rm as} \qquad r\to
\infty}}\ .}
It is probably possible to check this relation by calculating
numerically the correlation function\ \twopt\ with 
the use of exact
form-factors\ \Fedya. It is also worth noting that the expression
\main\  is
expected to hold for 
the one-point functions of the sinh-Gordon model
as well, if one makes the
substitution $\beta^2\to -\beta^2$ in\ \main.
The form-factors for the
sinh-Gordon model can be found in\ \Fedya,\ \KouM.

The one-point functions\  \expect\ 
of the  sine-Gordon model can be used
to derive the one-point
functions of primary fields in $c<1$ ``minimal''
CFT perturbed with the operator $\Phi_{1,3}$.
As is known, the perturbed
``minimal model'' ${\cal M}_{p/p'}$
(with $c =1-{6(p'-p)^2\over p\, p'}$) can be obtained by 
``quantum group restrictions''
from the sine-Gordon QFT\ \sg\ with 
$\xi={p\over p'-p}$\ \Leclair,\ \Sm.
Using this relation one obtains from \ \main\ 
\eqn\philk{\langle\,\Phi_{l,k}\, \rangle =
\biggl[\, M\ {{\sqrt{\pi}\, \Gamma\big(
{3\over 2}+{\xi\over 2}\big)}\over{2\
\Gamma\big({\xi\over 2}\big)}}
\, \biggr]^{2\Delta_{l,k}}\ {\cal Q}\big((\xi+1)l-\xi k\big)\ ,}
where the function
$\cal Q(\eta)$ for $\big|\,\Re e\ \eta\, \big|<\xi$
is given by the
integral 
$${\cal Q}(\eta)={\rm exp}\biggl\{
\int_{0}^{\infty} {d t\over t}\ \bigg(\,
{  {\rm cosh}(2 t)\  {\rm sinh}\big( t (\eta-1)\big)\ 
{\rm sinh}\big( t (\eta+1)\big)
\over 2\,  {\rm cosh}(t)\,
{\rm sinh}( t \xi )\,   {\rm sinh}\big(t(1+\xi)
\big)}-
{ (\eta^2-1 )\over 2  \xi (\xi+1)}\,  e^{-4t}\, \bigg)\biggr\} $$ 
and it is defined 
through analytic continuation outside this domain.
In\ \philk\ $\Phi_{l,k}$ stands
for the primary field of the dimension
\eqn\deltalk{
\Delta_{l,k} = {{\big((\xi+1)l-\xi k\big)^2-1}\over{4\xi(\xi+1)}}\ }
with canonical normalization
\eqn\phinorm{
\langle\,
\Phi_{l,k}(x)\ \Phi_{l,k}(y)\, \rangle
\to |x-y|^{-4\Delta_{l,k}}\qquad {\rm as}
\qquad |x-y|\to 0\ ,}
and $M$ denotes physical mass in the perturbed theory.

Finally, we remark that accepting the
conjecture\ \main\ one can make 
an interesting prediction
about long distance asymptotic of two-point 
correlation function in the  XXZ spin chain
\eqn\xxzass{
{{\langle\, vac\, |\, 
\sigma_{s}^{x}\, \sigma_{s+n}^{x}\, |\, vac\,
\rangle }\over{\langle\, vac\, |\, vac \, \rangle }}
\to F(\beta^2)\,\ n^{-\beta^2}\quad {\rm as}
\quad n\to\infty\ ,}
where $|\, vac\,
\rangle $ is the ground state of the  XXZ Hamiltonian \foot{As usual,
the infinite XXZ spin chain is defined as $N\to\infty$ limit of a
finite chain of size $N$ with periodic boundary conditions.}
\eqn\xxz{{\bf H}_{XXZ}=-{{1-\beta^2}\over{2\,\sin(\pi\beta^2)}}
\sum_{s=-\infty}^{\infty}\big(
\sigma_{s}^{x}
\sigma_{s+1}^{x}+\sigma_{s}^{y}\sigma_{s+1}^{y}+\cos(\pi\beta^2)
(\sigma_{s}^{z}\sigma_{s+1}^{z}-1)\big)\ ,}
$\sigma_{s}^{x}, \sigma_{s}^{y}$ and $\sigma_{s}^{z}$
are Pauli matrices
associated with the site $s$ of the chain. While the power-like
behavior in\  \xxzass\  is well known, the factor
\eqn\xxzfactor{\eqalign{F(\beta^2)=&{(1+\xi)^2\over 2}\ 
\biggl[\, { \Gamma\big({\xi\over 2}\big)\over
2\sqrt\pi\  \Gamma\big({1\over 2}+{\xi\over 2}\big)}\,
\biggr]^{\beta^2}\times\cr & 
{\rm exp}\biggl\{-
\int_{0}^{\infty} {d t\over t}\ \Big(\,
{ {\rm sinh}\big(\beta^2 t  \big)\over
{\rm sinh}( t)\, 
{\rm cosh}\big((1-\beta^2) t\big )}-
\beta^2\,   e^{-2 t}\, \Big)\biggr\} }}
in\ \xxzass\ is the consequence of\ \main\ (see Appendix D).

\hskip2.0cm

\centerline{\bf Acknowledgments}

\hskip0.5cm

S.L. acknowledges stimulating  discussions with A. LeClair.
The work of S.L.
is supported in part by NSF grant. A.Z. is grateful to
Al. Zamolodchikov
for interest to this work and remarks. The  research of A.Z.
is supported by DOE grant \#DE-FG05-90ER40559.

\vskip 0.4in

\appendix{A}{}
\vskip 0.2in

The expectation value\ \expect\ can be calculated as the Euclidean 
functional integral
\eqn\fint{{\cal G}_a=Z^{-1}(\mu)\,\int\big[{\cal
D}\varphi\big]\,e^{ia\varphi(0)}\,e^{-{\cal A}_{SG}}\ ,}
where ${\cal A}_{SG}$ is the action\  \sg. For $a=\alpha/\beta$,
$\beta\to 0$ this integral is dominated by a saddle-point 
configuration $\varphi(x) = \phi(x)/\beta$ which solves the 
classical equation
\eqn\classeq{
\partial_{\nu}^2\,
\phi(x) = m^2 \,\sin\phi(x) - 8\pi i \alpha\,  \delta^{(2)}(x)\ ,}
and decays sufficiently fast at $|x|\to\infty$. Obviously, the solution
depends only on the radial coordinate  $\phi = \phi(r), \ 
r=|x|$. Note that
$\phi$ is real only if $\alpha$ is imaginary,
$\alpha=i\omega$. The classical action calculated on this solution
diverges. Proper treatment of this singularity requires introducing
small cutoff distance $\varepsilon$ and taking the limit
\eqn\classac{
\log \langle\,e^{-{\omega\over\beta}\varphi(0)}\, \rangle
\sim
{1\over\beta^2}\ 
\lim_{\varepsilon\to 0}\,\big(\, {\omega^2}\, \log(\varepsilon^2)
-{\omega}\, \phi(\varepsilon)-{2}\, S(\omega,\varepsilon)\, \big)\ ,} 
where 
\eqn\radac{
S(\omega,\varepsilon)=
\int_{\varepsilon}^{\infty}\, {rdr\over 16}\ 
\biggl\{(\partial_r \phi)^2 + 2m^2\, (1-\cos\phi)\biggr\}\ ,}
and $\phi(r)$ is the solution to the equation
\eqn\radcl{
\partial_{r}^2 \phi +  r^{-1}\, \partial_{r}\phi -
m^2 \sin\phi = 0\ ,}
which satisfies the asymptotic conditions
\eqn\classass{
\phi(r)\to 4\,\omega\,\bigg[\,\log\Big({{mr}\over 2}\Big)+
C_{\omega}\, \bigg] \quad {\rm as} \quad r\to 0\ ,}
\eqn\classass{
\phi(r)\to -{4\over\pi}\, \sinh(\pi\omega)\ K_{0}(mr)
\quad {\rm as} \quad
r\to\infty\ .}
Here $ C_{\omega}$ is certain constant and $K_0 (t)$ is the
Macdonald function. Let us introduce the function
\eqn\somega{ S(\omega) = \lim_{\varepsilon\to 0}\bigg(\omega^2
\log\Big({{m\varepsilon}\over 2}\Big)+
S(\omega,\varepsilon)\bigg)\ ,}
Taking derivative of\ \classac\ with
respect to $\omega$
and using the equation\ \radcl\ one can derive the relation
\eqn\logo{
\log \langle\,e^{-{\omega\over\beta}\varphi(0)}
\, \rangle \sim {2\over\beta^2}\bigg(
-\omega^2
\log\Big({m\over 2}\Big)+
\omega\,\int_{0}^{\omega}{{d\tau}\over{\tau^2}}
\ S(\tau)\bigg)\ .}
As is shown in Appendix B
the function\ \somega\ admits the following
representation
\eqn\spsi{S(\omega)=\omega^2 - \int_{0}^{\infty} 
\, {{d\nu}\over{2\pi}}\ 
{\rm log }\biggl( {{\rm cosh}^2\pi(\nu-\omega)\,
{\rm cosh}^2\pi(\nu+\omega)
\over {\rm cosh}^2\pi\nu\ {\rm cosh}\pi(\nu-2\omega)\, 
{\rm cosh}\pi(\nu+2\omega)} \biggr)\
\Re e\, \Psi\big({1\over 2}-i\nu\big)\ ,}
where $\Psi(t)=\Gamma'(t)/\Gamma(t)$.
The exponential factor in\ \expcl\ 
is derived from\ \logo\ and \spsi\ with the use of the integral
representation for $\Psi(t)$. The pre-exponential factor
in\ \expcl\ can 
be obtained by evaluating the functional integral\ \fint\ in
the Gaussian approximation
around the above classical solution $\phi$
\eqn\det{D(i\omega)=
\bigg({{m\varepsilon}\over 2}\,e^{\gamma}\bigg)^{2\omega^2}\,
\bigg[\, 
{\rm Det}'\bigg({{-\partial_{\mu}^2 + m^2
\cos\phi}\over{
-\partial_{\mu}^2 + m^2}}\bigg)\, \bigg]^{-{1\over 2}}\ ,} 
where\ $\gamma = 0.577216...$\ is the Euler constant.
The first factor in\ \det\ appeared because of the mass
renormalization in \logo; its main role is to cancel ultraviolet
divergences
in the determinant. Note that validity of\ \main\ requires
the determinant\ \det\ to be
\eqn\ddet{D(i\omega)=
2^{2\omega^2}\,\exp\bigg(-{1\over 2}\int_{0}^{\infty}
{{dt}\over t}\,{{\sin^2 (2\,\omega\, t)}\over{\cosh^2 t}}
\bigg)\ .}
It would be very interesting to evaluate the
functional determinant\  \det\ directly
and check\ \ddet.

\vskip 0.2in

\appendix{B}{}

\vskip 0.2in

The one-point function ${\langle \,
e^{ia\varphi}\,  \rangle}$ can be expressed in terms
of the appropriately regularized (see below)
Euclidean functional integral
\eqn\ffint{
{\cal I}(a)=\int_{{\cal F}_a}\big[{\cal D}\psi
{\cal D}{\bar\psi}\big]\,
e^{-{\cal A}_{MTM}}\ ,}
where ${\cal A}_{MTM}$ is the Euclidean version of
the action\ \mtm\ and
the integration is taken over the space ${\cal F}_a$ of ``twisted''
field configurations,
such that $\psi(x)$ and ${\bar\psi}(x)$ acquire
phases
\eqn\monod{\psi\to e^{i{2\pi a\over \beta}}\, \psi ,
\qquad {\bar\psi} \to e^{-i{2\pi a\over \beta}}\, {\bar\psi}\ ,}
when continued around the point $x=0$
\foot{In other words,
the fields $\psi, {\bar\psi}$ are defined on the universal cover of the
punctured Euclidean
plane ${\bf R}^2/\lbrace 0 \rbrace$ and satisfy
there the quasi-periodicity
condition\ \monod.}. In general case it is
not known how one can evaluate the functional
integral\ \ffint\ directly.
In the case $\beta^2 = 1/2$ the action\ \mtm\ becomes
quadratic and the
task simplifies drastically.

The problem is most conveniently
treated in conformal polar coordinates
$(\eta,\theta)$
\eqn\polar{
z=x+iy = e^{\eta+i\theta}\ , \quad {\bar z} =
x-iy = e^{\eta-i\theta}\ .}
In this coordinates the action\ \mtm\ with $g=0$ takes the form
\eqn\ffact{
{\cal A}_{FF} =i\int_{0}^{2\pi}
d\theta\int_{-\infty}^{\infty}d\eta\,
\Big\{\Psi_{L}^{\dagger}
(\partial_{\theta}-i\partial_{\eta})\Psi_{L} +
\Psi_{R}^{\dagger}
(\partial_{\theta}+i\partial_{\eta})\Psi_{R} - i M\,
e^{\eta}\,
\big(\Psi_{L}^{\dagger}\Psi_{R} +
\Psi_{L}\Psi_{R}^{\dagger}\big)\Big\}\, ,}
where $\Psi_{L,R}(\eta,\theta)$ and
$\Psi_{L,R}^{\dagger}(\eta,\theta)$
are the components of the Dirac bi-spinors $\psi$ and ${\bar\psi}$
transformed to the coordinates\ \polar\foot{
$\psi_{L}(x,y)=
e^{-{1\over 2}\eta-
{i\over 2}\theta}\Psi_{L}(\eta,\theta),\   \psi_{R}(x,y)=
e^{-{1\over 2}\eta + {i\over 2}\theta}\Psi_{R}(\eta,\theta)$, where
$\psi_{L,R}$ are the components of the bi-spinor $\psi$ in Cartesian
coordinates $x,y$, and
similarly for $\Psi^{\dagger}$.}.
According to \monod\ they
satisfy the quasi-periodicity conditions
\eqn\quasiper{
\Psi_{L,R}(\eta, \theta+2\pi) = - e^{2\sqrt{2}\pi ia}\,
\Psi_{L,R}(\eta, \theta),
\qquad \Psi_{L,R}^{\dagger}(\eta, \theta+2\pi) =
- e^{-2\sqrt{2}\pi ia}\,
\Psi_{L,R}(\eta, \theta)\ .}

As usual, the most efficient way to evaluate the
functional integral\ \ffint\ is to use the
Hamiltonian formalism. The Hamiltonian picture which
appears most natural in our case is the
one where the polar angle $\theta$
is treated as (Euclidean) time,
and the Hilbert space ${\cal H}$
is associated
with the ``equal time'' slice $\theta=const$. We call this approach
{\it angular quantization}\foot{Angular quantization can be regarded
as QFT version of Baxter's
Corner Transfer Matrix approach\ \Baxter.
Angular quantization was previously used in\ \Luk\ in calculation
the sine-Gordon form-factors and correlation functions of Jost
operators.}. In this Hamiltonian formalism the fields
$\Psi, \Psi^{\dagger}$ become operators acting in ${\cal H}$.
They
satisfy canonical equal-time anti-commutation relations
\eqn\cancomm{\{\Psi_{L}(\eta), \Psi_{L}^{\dagger}(\eta')\} =
\delta(\eta-\eta')\ ,
\qquad \{\Psi_{R}(\eta), \Psi_{R}^{\dagger}(\eta')\} =
\delta(\eta-\eta')\ ,}
while the angular Hamiltonian derived from\ \ffact\ has the form
\eqn\ffham{
{\bf K}=i\int_{-\infty}^{\infty}{d\eta}\,
\Big\{- \Psi_{L}^{\dagger}\partial_{\eta}
\Psi_{L} + \Psi_{R}^{\dagger}\partial_{\eta} \Psi_{R} - M\,e^{\eta}
\,\big(\Psi_{L}^{\dagger}\Psi_{R} +
\Psi_{L}\Psi_{R}^{\dagger}\big)\Big\}\ .}
The $\eta$-dependence of the mass
term here has the effect of preventing the
fermions from penetrating too
far in the positive $\eta$ direction. We
call this effect the ``mass barrier''.
Correspondingly, the Hamiltonian\ \ffham\ is
diagonalized by the following
decompositions
\eqn\decomp{\eqalign{&
\Psi_{L}(\eta,\theta)=\int_{-\infty}^{\infty}\,
{d\nu\over \sqrt{2 \pi}}\ 
c_{\nu}\ u_{\nu}(\eta)\, e^{-\nu \theta}\ , \qquad
\Psi_{R}(\eta,\theta) 
=\int_{-\infty}^{\infty}\,
{d\nu\over \sqrt{2 \pi}}\ 
c_{\nu}\
v_{\nu}(\eta)\, e^{-\nu\theta}\ , \cr
&\Psi_{L}^{\dagger}(\eta,\theta) =
\int_{-\infty}^{\infty}\,
{d\nu\over \sqrt{2 \pi}}\ 
c_{\nu}^{\dagger}\
u_{\nu}^{*}(\eta)\, e^{\nu\theta}\ ,
\qquad \Psi_{R}(\eta,\theta)
=\int_{-\infty}^{\infty}\,
{d\nu\over \sqrt{2 \pi}}\ 
c_{\nu}^{\dagger}\
v_{\nu}^{*}(\eta)\, e^{\nu\theta} }}
in terms of the partial waves
\eqn\waves{
\biggl({{u_{\nu}(\eta)}\atop{v_{\nu}(\eta)}}\biggr)=
{\sqrt{2M}\
e^{\eta\over 2} \over \Gamma\big({1\over 2}-i\nu\big)}\,
\bigg({M\over 2}\bigg)^{-i\nu}\
\biggl({{K_{\ {1\over 2}-i\nu}\big(Me^{\eta}\big)\atop{K_{{1\over
2}+i\nu}\big(Me^{\eta}\big)}}}\biggr)\ ,}
which describe the fermion scattering off the ``mass barrier''
\eqn\fass{
\bigg({{u_{\nu}(\eta)}\atop{v_{\nu}(\eta)}}\bigg)\to
\bigg({1\atop 0}\bigg)\,e^{i\nu\eta} +
S_{F}(\nu)\,\bigg({0\atop 1}\bigg)\,e^{-i\nu\eta}
\quad {\rm  as} \quad \eta\to -\infty\ .}
Here
\eqn\sf{S_{F}(\nu) =\bigg({M\over 2}\bigg)^{-2i\nu}\
{{\Gamma\big({1\over 2}+i\nu\big)}
\over{\Gamma\big({1\over 2}-i\nu\big)}}  }
is the S-matrix associated with this process.
In\ \decomp\ $c_{\nu}$ and
$c_{\nu}^{\dagger}$ are operators
satisfying the anti-commutation
relations
\eqn\kliff{\eqalign{
&\{c_{\nu}, c_{\nu'}\} =
\{c_{\nu}^{\dagger}, c_{\nu'}^{\dagger}\} = 0\ ,\cr
&\{c_{\nu}, c_{\nu'}^{\dagger}\} =
\delta(\nu-\nu')\ .}}
The Hilbert
space ${\cal H}$ is the
fermionic Fock space over the
algebra\ \kliff\ with the vacuum state $|\, v \rangle$
which satisfies
the equations
\eqn\fockvac{\eqalign{
&c_{\nu}\,|\, v \rangle = 0 \qquad {\rm for} \qquad \nu > 0\ ,\cr
&c_{\nu}^{\dagger}\, |\, v \rangle = 0 \qquad
{\rm for} \qquad \nu < 0\ .}}
The angular  Hamiltonian\ \ffham\ can be written as
\eqn\aaham{
{\bf  K}=\int_{0}^{\infty}
d\nu\, \nu
\, \big(\, c_{\nu}^{\dagger} c_{\nu} + c_{-\nu}
c_{-\nu}^{\dagger}\, \big)\ .}

As the ``Euclidean time'' $\theta$ in\ \ffact\ is
compactified the functional
integral\ \ffint\ (with $g=0$) is given by the trace
\eqn\trace{
{\cal I}(a) =
{\rm Tr}_{\cal H}\,\biggl[\,
e^{-{2\pi  {\bf K}}+i{2\pi a\over \beta}\,
{\bf Q}}\, \biggr]\ ,}
where the the fermion charge
\eqn\charge{
{\bf Q} = \int_{-\infty}^{\infty}
d\eta\ \big(\, \Psi_{L}^{\dagger}\Psi_{L}
+\Psi_{R}^{\dagger}\Psi_{R}\, \big) =
\int_{0}^{\infty}d\nu
\  \big(\, c_{\nu}^{\dagger} c_{\nu} + c_{-\nu}
c_{-\nu}^{\dagger}\, \big) }
is introduced in order to
impose the twisted boundary conditions\ \monod.

The trace\ \trace\ requires
regularization. One can
define the system\ \ffact\ as
the limiting case $\varepsilon \to 0$ of similar system in a
semi-infinite box $\eta\in\big[\, \log\varepsilon,
\infty\, \big)$\foot{This
corresponds to cutting out a small disc
of the size $\varepsilon$ around 
the puncture $x=0$ in the functional
integral\ \ffint.
The other infinity $\eta\to +\infty $ does not cause any
problem because of the
``mass barrier''.} with the boundary conditions
\eqn\bc{
\big[\, \Psi_{L}(\eta)-\Psi_{R}(\eta)
\, \big]_{\eta = \log\varepsilon}=
\big[\, \Psi_{L}^{\dagger}(\eta)-
\Psi_{R}^{\dagger}(\eta)\, \big]_{\eta =
\log\varepsilon} = 0\ .}
Let us denote
${\cal I}_{\varepsilon}(a)$ the
trace\ \trace\ regularized this
way. Simple analysis
(which takes into account the form of the boundary
state associated with the boundary
conditions\ \bc\ in conformal case
$M=0$) shows that
\eqn\ieps{
\langle\, e^{ia\varphi}\,  \rangle =
\lim_{\varepsilon\to 0}\,
\varepsilon^{-2 a^2}\,
{{\cal I}_{\varepsilon}(a)}/
{{\cal I}_{\varepsilon}(0)}\ .}

Using the above regularization
one can directly evaluate\ \trace, \ieps,
with the result\nobreak\foot{This result was also obtained by
Al. Zamolodchikov\ \AlZ.}
\eqn\kdidu{
\langle\,
e^{ia\varphi}\,  \rangle
=\exp\biggl\lbrace
\int_{0}^{\infty}{{d\nu}\,
\over{2\pi i}}\,\log
\bigg({\big(1+e^{-2\pi\nu+2\sqrt{2}i\pi a}\big)
\big(1+e^{-2\pi\nu-2\sqrt{2}i\pi a}\big)\over
\big(1+e^{-2\pi\nu}\big)^2}\bigg)\,
\partial_{\nu}\log S_{F}(\nu)
\biggl\rbrace\ ,}
where $S_{F}(\nu)$ is the S-matrix\ \sf.
Using the integral
representation
\eqn\sfint{
{1\over i}\,\log S_{F}(\nu) =-2\nu\ 
{\rm log}\bigg({M\over 2}\bigg)-\int_{0}^{\infty}{{dt}\over t}\ 
\bigg[{{\sin (2\nu t)}\over{\sinh t}}-2\nu\,e^{-2t}\bigg]\ }
one arrives at\ \expff.

Eq.\ \expff\ can be checked against exact result of\ \WuM\ for the 
spontaneous magnetization in
the  Ising model. Indeed, at the free-fermion 
point the operator
$
{\rm cos} \big({\beta\over 2} \varphi\big)$
can be expressed in term of two independent Ising spin fields
$\sigma^{(i)}(x)\ (i=1,2)$  as
\eqn\liuuy{{\rm cos}\Big({\beta\over 2} \varphi(x)\Big)
\Big|_{
\beta^2={1\over 2}}=
2^{-{1\over 2}}\ \sigma^{(1)}(x)\,  \sigma^{(2)}(x)\ ,}
where $\sigma^{(i)}(x)$ are normalized as follows
\eqn\nir{ \sigma^{(i)}(x)\, 
 \sigma^{(i)}(0)\to |x|^{-1/4}\ , \ \ \ |x|\to 0\ .}
According to\ \WuM
\eqn\mcacc{\langle\, \sigma(0)\, \rangle^2=M^{{1\over 4}}\ 
2^{{1\over 6 }}\,  e^{-{1\over 4}}
\ A^3\ ,}
where $A$ is the Glaisher constant. This constant can be expressed in 
term of the derivative of the Riemann zeta function,
\eqn\azeta{A=
{\rm exp}\Big(-\zeta'(-1)+{1\over 12}\Big)=1.28242712910062...\ .}
Evaluating the integral 
in\ \expff\ for\ $a= 2^{-{3\over 2}}$ and using \ \azeta\ one finds
complete agreement with\ \mcacc.

It is instructive to
compare our one-point function\ \expff\ with known
exact results for
the two-point function\ \twopt\ in
the free-fermion theory\ \MiwJ,\ \BerLec. For $\beta^2=1/2$
and $aa'>0$\ \twopt\ can be written as
\eqn\twoo{
\langle\,
e^{ia\varphi}(x)\, e^{ia'\varphi}(0)\,  \rangle
=\langle\,
e^{ia\varphi}\,  \rangle\langle\,e^{ia'\varphi}\,  \rangle\ 
\exp\Big(\Sigma_{aa'}\big(M|x|/2\big)\Big)\ ,}
with
$\Sigma_{aa'}(\tau)$ denoting the integral
\eqn\sdef{\Sigma_{aa'}(\tau)=-{1\over2} \int_{\tau}^{\infty}\,
\rho d\rho\, \bigg\{\, 
(\partial_{\rho}\chi)^2 - 4\sinh^2 \chi -
{2\, {(a-a')^2}\over{\rho^2}}\tanh\chi\, \bigg\}\ .}
Here $\chi(\rho)$ is the solution of the differential equation
\eqn\differ{\partial^2_{\rho}\chi+\rho^{-1}\partial_{\rho}\chi= 
2\, {\rm sinh}(2 \chi)+{2\,
(a-a')^2\over \rho^2}\ {\rm tanh }\,\chi\, 
\big(\, 1-{\rm tanh }^2\chi\, \big),}
subject to the asymptotic conditions
\eqn\penass{\eqalign{
&\chi(\rho)\to \sqrt{2}\, (a+a')\ \Big[\,  
{\rm log}
(\rho)+C_{aa'}\, \Big]\ ,\qquad {\rm as} \qquad \rho\to 0\ ,\cr
&\chi(\rho)\to
 -{2\over
\pi}\,  \Big(\, {\rm sin}(\sqrt{2}\pi\, a) \, {\rm sin}
({\sqrt 2}\pi\,  a')\, \Big)^{{1\over 2}}
\ K_{\sqrt{2}(a-a')}(2\rho)\ ,
\qquad {\rm  as} \qquad \rho\to\infty\ ,}}
and $C_{aa'}$ is a  certain constant.
Now, one can take the limit $|x|\to
0$ in\ \twoo\ and use\ \ope\  to obtain the relation
\eqn\slm{
\Sigma_{a a'}\equiv\lim_{\tau\to 0}\Big(-4aa'\log\tau+
\Sigma_{aa'}(\tau)\, \Big)
=\log\bigg(\, \Big({M\over 2}\Big)^{-4 a a'}\,  
{{{\cal G}_{a+a'} }
\over{{\cal G}_{a}\,{\cal G}_{a'}}}\, \bigg)\ , }
where ${\cal G}_a$ is the one-point function\ \kdidu\ and
$$ a a'>0, \ \ \ \ |a+a'|< 2^{-{3\over 2}}\ .$$
We have checked the relation\ \slm\ numerically.
Note that for
pure imaginary $a=i\,  \omega/\sqrt{2}$,
$a'=i\, \omega'/\sqrt{2}$ the
function\ $\chi(\rho)$ also
becomes pure imaginary.
If in addition $\omega'=
\omega$, then   
$$\phi(r)=-2i\,
\chi\big(mr/2\big) $$
satisfies\ \radcl,\ \classass\ in Appendix A, and hence
\eqn\lsksj{
S(\omega)={1\over 2}\, 
\Big(\, \Sigma_{aa'}
-4 \int_0^{\infty}d\rho\ \rho\, \sinh^2\chi\ 
\Big)\Big|_{a=a'=i\, \omega/\sqrt{2}}\  ,}
where\ $S(\omega)$\ is  defined by\ \somega
\ . 
The integral in\ \lsksj\ can be evaluated by
observing that
\eqn\msns{4 \rho\, \sinh^2\chi={1\over 2}\ \partial_{\rho} 
\Big[\, \rho^2\,  \big(\, 
4 \sinh^2\chi-(\partial_{\rho}\chi)^2\, \big)\, \Big]\ .}
Using\ \slm,\ \kdidu\ one arrives at\ \spsi.

\vskip 0.2in

\appendix{C}{}

\vskip 0.2in

Expanding \sigman\ in power series in $\beta^2$ one
finds
\eqn\saa{\sigma_2 = - 4\, \big(\gamma + \log(m/2)\big) +
\big(
\zeta(3)-\pi^2\big)\, {\beta^4\over 3} + O(\beta^6)\ ,}
\eqn\saaaa{\sigma_4 = 56\,
\zeta(3)\,\beta^2\ +\ 72\,\zeta(3)\,\beta^4\ +\ O(\beta^6)\ ,}
where\ $\zeta(s)$ is the Riemann zeta function and $\gamma$ is the 
Euler constant.
These expansions are to be compared with the results of 
perturbation theory for the action\ \sg. In the
perturbative calculations it is more convenient
to use another 
(equivalent to\ \sigmadef)
definition of the ``fully connected''
one-point functions $\sigma_{2n}$
\eqn\sgaa{\sigma_2 = \langle\,\varphi^2\, \rangle
=\lim_{\varepsilon\to
0}\big(\, \langle\, \varphi(x)\varphi(0)\, \rangle
\big|_{|x|=\varepsilon}+4\log\varepsilon \, \big)\ ,}
\eqn\sgaaaa{
\sigma_{2n}= \langle\,
\varphi(x_1)\varphi(x_2) \ldots
\varphi(x_{2n})\,
 \rangle_{c}\big|_{x_1=x_2=\ldots=x_{2n}}\ ,}
where $\langle\ldots\rangle_c$ 
in\ \sgaaaa\ is {\it connected} $2n$-point correlation
function. The Feynman diagrams contributing to
$\sigma_2$ and $\sigma_4$ up to the
order $\beta^4$ are shown in Fig.1.
The calculations are
straightforward\foot{One only has to be
sure to express the result in terms of the physical mass $m$.} 
and the result coincides with\ \saa,\ \saaaa.

Expansion of\ $\sigma_2$\ \sigman\ around $\beta^2=1/2$ has the form
\eqn\stg{
\sigma_2 = -4\,\big(1+\gamma+\log(M/2)\big)+{g\over\pi}\,
\big(7\, \zeta(3)-2 \big)
+ O(g^2)\ ,}
where $g=\pi\,(1/2-\beta^2)/\beta^2$. The expectation value
$\langle\,
\varphi^2\, \rangle$
can be calculated in perturbation theory of the Massive
Thirring model\ \mtm\ since \jphi\ allows one to relate
the two-point function $\langle\,
\varphi(x)\varphi(y)\, \rangle$
in\ \sgaa\ to the 
current-current correlation function
\eqn\jj{
\langle\, 
j^{\mu}(x)\,j^{\nu}(0)\, \rangle = \int {{d^2 k}\over
{(2\pi)^2}}\ \Big(\delta^{\mu\nu}-
{{k^{\mu}k^{\nu}}\over k^2}\Big)\, \Pi
(k^2)\,e^{ikx}\ ,}
\eqn\pp{{\big({{\beta}/{2\pi}}\big)^2}
\, \langle\, \varphi(x)\varphi(0) \, \rangle
= \int
{{d^2k}\over{(2\pi)^2}}\,\Pi(k^2)\,e^{ikx}\ .}
The first two terms of the perturbative expansion in $g$ for the
function $\Pi(k^2)$ can be obtained by calculating the contributions
of the Feynman  diagrams in Fig.2
\eqn\pipert{
M^2\, \Pi(k^2)=
{1\over{4\pi}}\,{{\sinh \theta - \theta}\over{\sinh\theta\,
\sinh^2 (\theta/2)}} -
{g\over{4\pi^2}}\,\bigg[{{(\sinh\theta-\theta)^2}\over{\sinh^2 \theta\,
\sinh^2 (\theta/2)}}-{{\theta^2}\over{\sinh^2\theta}}\bigg]
+ O(g^2)\ ,}
where $\theta$ is related to $k^2$ as
$$
k^2 = 4\,M^2\,\sinh^2(\theta/2)\ .
$$
Evaluating the Fourier transform\ \pp\ and taking
the limit\ \sgaa\ one
obtains exactly\ \stg.

\vskip 0.2in

\appendix{D}{}

\vskip 0.2in

Exact integral representation for the correlation function  
\eqn\cn{{\langle\, 
{vac\, |\, \sigma_{s}^{x}\sigma_{s+n}^{x}\,
|\, vac\, \rangle}\over{\langle\, 
vac\, |\, vac\, \rangle}}\ }
was found recently by Jimbo and  Miwa\ \MiW.
However in this representation the
correlation function\ \cn\ is
expressed in terms of $n$-fold integral and
therefore finding $n\to\infty$ asymptotic form of \cn\ remains a 
challenging problem.
Here we will explain how the asymptotic 
formula\ \xxzass,\ \xxzfactor\ follows
from our conjecture \main.

The XXZ model is a limiting case of XYZ spin chain 
\eqn\xyz{
{\bf H}_{XYZ}=-{1\over{2\varepsilon}}\sum_{s=-\infty}^{\infty}
\big(J_{x}\,\sigma_{s}^{x}
\, \sigma_{s+1}^{x}+
J_{y}\,\sigma_{s}^{y}
\, \sigma_{s+1}^{y}+J_{z}\,\sigma_{s}^{z}
\, \sigma_{s+1}^{z}-J\big) \ ,}
with $J_{x}\geq J_{y}\geq |J_{z}|$; we introduced here an auxiliary
parameter $\varepsilon$ which
will be interpreted as a lattice spacing.
Following Baxter\ \Baxter, we use the parameterization
\eqn\psosi{\eqalign{
&J_x={1-\beta^2\over \pi }\biggl(
{\theta_{4}(\beta^2) \theta'_{1}(0)\over
\theta_{4}(\beta^2) \theta_{1}(\beta^2)}+
{\theta_{1}(\beta^2) \theta'_{1}(0)\over
\theta_{4}(\beta^2) \theta_{4}(\beta^2)} \biggr)\ ,\cr
&J_y={1-\beta^2\over \pi }\biggl(
{\theta_{4}(\beta^2) \theta'_{1}(0)\over
\theta_{4}(\beta^2) \theta_{1}(\beta^2)}-
{\theta_{1}(\beta^2) \theta'_{1}(0)\over
\theta_{4}(\beta^2) \theta_{4}(\beta^2)} \biggr)\ ,\cr
&J_z={1-\beta^2\over \pi }\biggl(
{\theta'_{1}(\beta^2)\over \theta_{1}(\beta^2)}-
{\theta'_{4}(\beta^2)\over \theta_{4}(\beta^2)}  \biggr)\ ,\cr
&J=-{1-\beta^2\over \pi }\biggl(
{\theta'_{1}(\beta^2)\over \theta_{1}(\beta^2)}+
{\theta'_{4}(\beta^2)\over \theta_{4}(\beta^2)}  \biggr)\ . }}
Here
$$\eqalign{&\theta_{1}(v)=2 p^{{1\over 4}}\   {\rm sin}(\pi v)\ 
\prod_{n=1}^{ \infty} \big(1-p^{2 n}\big)\, \big(1-e^{2\pi i\, v}\,
p^{2 n}\big)\,
\big(1-e^{-2\pi i\, v}\,p^{2 n}\big)\ ,\cr 
&\theta_{4}(v)=
\prod_{n=1}^{ \infty} \big(1-p^{2 n}\big)\, \big(1-e^{2\pi i\, v}\,
p^{2 n-1}\big)\,
\big(1-e^{-2\pi i\, v}\,p^{2 n-1}\big)\ }$$
and the prime in\ \psosi\ means a  derivative. 
Note that for $p\to 0$ the difference 
$J_{x}-J_{y}$
vanishes and the  XYZ model reduces to the XXZ chain\ \xxz. 
As is known\ \luth, the XXZ 
model\ \xxz\ describes the critical behavior of XYZ chain, i.e.
the correlation length $R_c = \varepsilon N_c
\to \infty$ as $p\to 0$.
The scaling limit
$p\to 0,\ \   \varepsilon\sim p^{1\over{4 (1-\beta^2)}}\to
0$ of XYZ model is described by sine-Gordon QFT, and the following
relation between the operators holds
\eqn\sigmacos{
\sigma_{s}^{x} \to  N(\beta^2)\ \varepsilon^{\beta^2\over 2}\,
\cos\bigg({\beta\over 2}\varphi(r)\bigg)\, ,
 \qquad s\varepsilon\ \to r,}
where $N(\beta^2)$ is normalization factor.
The dependence of $n$ in \ \xxzass\ is
of course the simple consequence of
\sigmacos\ while the factor $
F$ in\ \xxzass\ is determined by the constant $N$, 
$F = {1\over 2}
N^2$. It is exact value of this constant which can be deduced
from\ \main\  with the use of two exact results in XYZ model:

a) Exact formula for
the energy gap due to Johnson, Krinski and McCoy\ \Jons,
which gives in the limit $p\to 0$ exact 
relation between $p$ and the
physical mass $M$ in the sine-Gordon model\ \luth
\eqn\Mq{
M = {4\over\varepsilon}\ p^{{1+\xi}\over 4}\ ,}
where the notation\ \xibeta\ is used. 

b) Exact result of Baxter and Kelland
for the expectation value of the spin
operator $\sigma_{s}^{x}$\ \BaxKel\ , which
in the limit $p\to 0$ reduces to
\eqn\expeq{{{\langle \, vac\, |\, \sigma_{s}^{x}\,
|\, vac\, \rangle} \over {\langle\,
vac\, |\, vac \rangle}}\  \to(1+\xi)\ p^{\xi\over 8}\ .}
Comparing\ \expeq\ with\ \main\ and
using\ \sigmacos, \Mq\ one derives \xxzfactor.

\listrefs

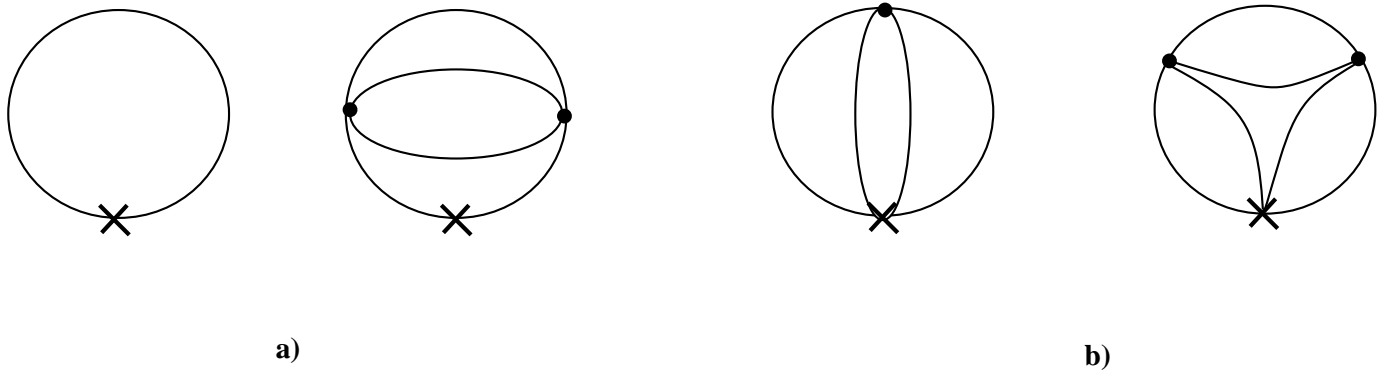
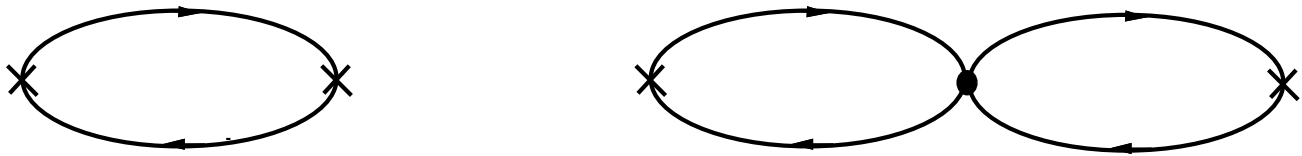
\end